\documentclass[12pt]{article}

\usepackage[noend,vlined,figure]{algorithm2e} 
\usepackage[utf8]{inputenc}
\usepackage{graphicx}
\usepackage{rotating}
\usepackage{tikz}
\usepackage{a4wide}

\newcommand{\circled}[1]{%
{\footnotesize%
\tikzstyle{every path} = [draw,ultra thick]%
\tikzstyle{every node} = [draw,shape=circle,very thick,minimum size=0.5cm,inner sep=0pt]%
\tikz[baseline,scale=0.71]{\node[anchor=base] (#1) {$#1$};}%
}}

\newcommand{\Construct}{\mbox{$\mathit{construct}$}}
\newcommand{\Minimum}{\mbox{$\mathit{minimum}$}}
\newcommand{\Insert}{\mbox{$\mathit{insert}$}}
\newcommand{\Extractmin}{\mbox{$\mathit{extract}$\textnormal{-}$\mathit{min}$}}

\newcommand{\Bulkinsert}{\mbox{$\mathit{bulk}${\rm -}$\mathit{insert}$}} 
\newcommand{\Siftdown}{\mbox{$\mathit{sift}$\mbox{\rm -}$\mathit{down}$}}
\newcommand{\Strengtheningsiftdown}{\mbox{$\mathit{strengthening}$\mbox{\rm -}$\mathit{sift}$\mbox{\rm -}$\mathit{down}$}}
\newcommand{\Rotatingsiftdown}{\mbox{$\mathit{swapping}$\mbox{\rm -}$\mathit{sift}$\mbox{\rm -}$\mathit{down}$}}
\newcommand{\Combinedsiftdown}{\mbox{$\mathit{combined}$\mbox{\rm -}$\mathit{sift}$\mbox{\rm -}$\mathit{down}$}}

\newcommand{\leafsearch}{\mbox{$\mathit{leaf}${\rm -}$\mathit{search}$}} 
\newcommand{\isleaf}{\mbox{$\mathit{is}${\rm -}$\mathit{leaf}$}} 
\newcommand{\leftchild}{\mbox{$\mathit{left}${\rm -}$\mathit{child}$}} 
\newcommand{\rightchild}{\mbox{$\mathit{right}${\rm -}$\mathit{child}$}} 
\newcommand{\parent}{\mbox{$\mathit{parent}$}} 
\newcommand{\sibling}{\mbox{$\mathit{sibling}$}}
  
\newcommand{\swapsubtrees}{\mbox{$\mathit{swap}${\rm -}$\mathit{subtrees}$}} 
\newcommand{\odd}{\mbox{$\mathit{odd}$}} 
\newcommand{\even}{\mbox{$\mathit{even}$}}
\newcommand{\bottomupsearch}{\mbox{$\mathit{bottom}${\rm -}$\mathit{up}${\rm -}$\mathit{search}$}} 
\newcommand{\interchange}{\mbox{$\mathit{interchange}$}} 
\newcommand{\shuffle}{\mbox{$\mathit{shuffle}$}}

\newtheorem{example}{Example}
 
\newtheorem{lemma}{Lemma} 

\title{Strengthened Lazy Heaps: Surpassing the Lower Bounds for Binary Heaps%
}

\author{%
Stefan Edelkamp \\
Faculty 3---Mathematics and Computer Science \\
University of Bremen \\
PO Box 330 440, 28334 Bremen, Germany
\and
Amr Elmasry \\
Department of Computer Engineering and Systems \\ 
Alexandria University \\
Alexandria 21544, Egypt 
\and
Jyrki Katajainen \\
Department of Computer Science \\ 
University of Copenhagen \\
Universitetsparken 5, 2100 Copenhagen East, Denmark
}

\begin{document}
\maketitle
\pagestyle{plain}
\pagenumbering{arabic}

\begin{abstract}
\noindent
Let $n$ denote the number of elements currently in a data structure.
An \emph{in-place} \emph{heap} is stored in
the first $n$ locations of an array, uses $O(1)$ extra space, and 
supports the operations: \Minimum{}, \Insert{}, and
\Extractmin{}.  We introduce an in-place heap, for which \Minimum{}
and \Insert{} take $O(1)$ worst-case time, and \Extractmin{} takes
$O(\lg{} n)$ worst-case time and involves at most $\lg{} n + O(1)$
element comparisons.  
The achieved bounds are optimal to within additive constant terms
for the number of element comparisons. In
particular, these bounds for both \Insert{} and \Extractmin{}---and
the time bound for \Insert{}---surpass the corresponding lower bounds
known for binary heaps, though our data structure is similar.  
In a binary heap, when viewed as a nearly complete
binary tree, every node other than the root obeys the heap
property, i.e.~the element at a node is not smaller than that at its
parent.  To surpass the lower bound for \Extractmin{}, we reinforce a
stronger property at the bottom levels of the heap that the element at
any right child is not smaller than that at its left sibling.  To
surpass the lower bound for \Insert{}, we buffer insertions and
allow $O(\lg^2 n)$ nodes to
violate heap order in relation to their parents.
\end{abstract}

\section{Introduction}

A \emph{binary heap}, invented by Williams \cite{Wil64}, is an
in-place data structure that 1)~implements a priority queue
(i.e.~supports the operations \Minimum{}, \Construct{}, \Insert{}, and
\Extractmin{}); 2)~requires $O(1)$ words of extra space in addition to
an array storing the current collection of elements; and 3)~is viewed
as a nearly complete binary tree where, for every node other than the
root, the element at that node is not smaller than the element at its
parent (\emph{min-heap property}).  Letting $n$ denote the number of
elements in the data structure, a binary heap supports \Minimum{} in
$O(1)$ worst-case time, and \Insert{} and \Extractmin{} in $O(\lg n)$
worst-case time.  For Williams' original proposal \cite{Wil64}, the
number of element comparisons performed by \Insert{} is at most $\lg n
+ O(1)$ and that by \Extractmin{} is at most $2 \lg n +
O(1)$. Immediately after the appearance of Williams' paper, Floyd
showed \cite{Flo64} how to support \Construct{}, which builds a heap
for $n$ elements, in $O(n)$ worst-case time with at most $2n$ element
comparisons.

Since a binary heap does not support all the operations
optimally, many attempts have been made to develop priority queues
supporting the same set (or even a larger set) of operations that
improve the worst-case running time of the operations as well as the
number of element comparisons performed by them
\cite{Bro78,Car91,CMP88,CEEK12,EEK13,EJK08,GM86,Vui78}.  In
Table~\ref{table:history} we summarize the fascinating history of the
problem, considering the space and comparison complexities.

When talking about optimality, we have to separate three different
concepts.  Assume that, for a problem of size $n$, the bound achieved
is $\mbox{A}(n)$ and the best possible bound is $\mbox{OPT}(n)$.
\begin{description}
\item[Asymptotic optimality:]  
  $\mbox{A}(n) = O\left(\mbox{OPT}(n)\right)$.
\item[Constant-factor optimality:]  
  $\mbox{A}(n) = \mbox{OPT}(n) + o\left(\mbox{OPT}(n)\right)$.
\item[Up-to-additive-constant optimality:]
  $\mbox{A}(n) = \mbox{OPT}(n) + O(1)$.
\end{description}

As to the amount of space used and the number of element comparisons
performed, we aim at up-to-additive-constant optimality.  From the
in\-for\-ma\-tion-theoretic lower bound for sorting \cite[Section
  5.3.1]{Knu73}, it follows that, in the worst case, either \Insert{}
or \Extractmin{} must perform at least $\lg n - O(1)$ element
comparisons. As to the running times, we aim at asymptotic optimality.
Our last natural goal is to support \Insert{} in $O(1)$ worst-case
time, because then \Construct{} can be trivially
realized in linear time by repeated insertions.

\begin{table}
\caption{\label{table:history}The worst-case comparison complexity of some
  priority queues. Here $n$ denotes the number of elements stored, $w$ is
  the size of machine words in bits, and the amount of extra space is
  measured in words. All these data structures support \Construct{} in
  $O(n)$ worst-case time and \Minimum{} in $O(1)$ worst-case time.}
\phantom{x}

\centering\begin{tabular}{|l|c|c|c|}
\hline
\multicolumn{1}{|c|}{\textbf{Data structure}\strut} & \textbf{Space} &
\Insert{} & \Extractmin{}\\ 
\hline
binary heaps \cite{Wil64}\strut & $O(1)$ & $\lg n + O(1)\strut$ & 
$2\lg n + O(1)$ \\
binomial queues \cite{Bro78,Vui78} & $O(n)$ & $O(1)$ & $2\lg n + O(1)$ \\
heaps on heaps \cite{GM86} & $O(1)$ & $\lg \lg n + O(1)\strut$ &  $\lg n + \log^{*} n + O(1)$ \\
queue of pennants \cite{CMP88} & $O(1)$ & $O(1)$ & $3\lg n + \log^{*} n + O(1)$ \\
multipartite priority queues \cite{EJK08} & $O(n)$ & $O(1)$ & $\lg n + O(1)$\\
engineered weak heaps \cite{EEK13} & $n/w + O(1)$ & $O(1)$ & $\lg n + O(1)$\\
strengthened lazy heaps [this work] & $O(1)$ & $O(1)$ & $\lg n + O(1)$ \\
\hline
\end{tabular}
\end{table} 

The binomial queue \cite{Vui78} was the first priority queue
supporting \Insert{} in $O(1)$ worst-case time. (This was mentioned as
a short note at the end of Brown's paper \cite{Bro78}.) 
 However, the binomial
queue is a pointer-based data structure requiring $O(n)$ pointers in
addition to the elements.  For binary heaps, Gonnet and Munro showed
\cite{GM86} how to perform \Insert{} using at most $\lg\lg n +O(1)$
element comparisons and \Extractmin{} using at most $\lg n + \lg^*n +
O(1)$ element comparisons. Carlsson et al.~showed \cite{CMP88} how to
achieve $O(1)$ worst-case time per \Insert{} by an in-place data
structure that utilizes a queue of pennants. (A pennant is a binary
heap with an extra root that has one child.) For this data structure,
the number of element comparisons performed per \Extractmin{} is
bounded by $3\lg n + \log^{*} n + O(1)$.  The multipartite priority
queue~\cite{EJK08} was the first priority queue achieving the
asymptotically-optimal time and additive-term-optimal comparison
bounds: $O(1)$ worst-case time for \Minimum{} and \Insert{}, and
$O(\lg n)$ worst-case time with at most $\lg n + O(1)$ element
comparisons for \Extractmin{}.  Unfortunately, the structure is
involved and its representation requires $O(n)$ pointers.  Another
solution by us~\cite{EEK13} is based on weak heaps~\cite{Dut93}.  To
implement \Insert{} in $O(1)$ worst-case time, we use a bulk-insertion
strategy---employing two buffers and incrementally merging one with
the weak heap before the other is full.  This priority queue also
achieves the desired worst-case time and comparison bounds, but it
uses $n$ additional bits, so the structure is still not quite
in-place.

For a long time, it was open  whether there exists an in-place data
structure that guarantees $O(1)$ worst-case time for \Minimum{} and
\Insert{}, and $O(\lg n)$ worst-case time for \Extractmin{} such that
\Extractmin{} performs at most $\lg n + O(1)$ element comparisons.  In
view of the lower bounds proved in \cite{GM86}, it was not entirely
clear if such a structure exists.  A strengthened lazy heap, 
proposed in this paper operates in-place and supports \Minimum{},
\Insert{}, and \Extractmin{} within the desired bounds, and thus
settles this long-standing open problem.

When a strengthened lazy heap is used in heapsort, the resulting algorithm sorts $n$
elements in-place in $O(n \lg{} n)$ worst-case time performing at most
$n \lg{} n + O(n)$ element comparisons. The number of element
comparisons performed matches the in\-for\-ma\-tion-theoretic lower
bound for sorting up to the additive linear term. Ultimate heapsort
\cite{ultimate} is known to have the same complexity bounds, but
the constant factor of the additive linear term is high.

In a binary heap the number of element moves performed by
\Extractmin{} is at most $\lg n + O(1)$. We have to avow that, in our
data structure, \Extractmin{} may require more element moves.  On the
positive side, we can adjust the number of element moves to be at most
$(1+\varepsilon) \lg n$, for any fixed constant $\varepsilon > 0$ and
large enough $n$, while still achieving the desired bounds for the
other operations.

One motivation for writing this paper was to show the limitation of
the lower bounds proved by Gonnet and Munro \cite{GM86} (see also
\cite{Car91}) in their important paper on binary heaps.  They showed
that $\lceil \lg \lg(n+2) \rceil - 2$ element comparisons are
necessary to insert an element into a binary heap.  In addition,
slightly correcting \cite{GM86}, Carlsson \cite{Car91} showed that
$\lceil \lg n \rceil + \delta(n)$ element comparisons are necessary
and sufficient to remove the minimum element from a binary heap that
has $n > 2^{h_{\delta(n)}+2}$ elements, where $h_1=1$ and $h_i =
h_{i-1} + 2^{h_{i-1}+i-1}$. One should notice that these lower bounds
are valid under the following assumptions:

\begin{enumerate}
\item[1)] All elements are stored in one nearly complete binary tree.
\item[2)] Every node obeys the heap property before and after each operation.
\item[3)] No order relation among the elements of the same level can be deduced from the element comparisons performed by previous operations.
\end{enumerate} 

We show that the number of element comparisons performed by
\Extractmin{} can be lowered to at most $\lg n +O(1)$ if we overrule
the third assumption by imposing an additional requirement that any
right child is not smaller than its left sibling; see Section
\ref{sec:strong}.  We also show that \Insert{} can be performed in
$O(1)$ worst-case time if we overrule the second assumption by
allowing $O(\lg^2 n)$ nodes to violate heap order; see Section
\ref{sec:buff}.  Lastly, we combine the two ideas and put everything
together in our final data structure; see Section
\ref{sec:ultimate}.

\section{Strong Heaps: Adding More Order}
\label{sec:strong}

A \emph{strong heap} is a binary heap with one additional invariant:
The element at any right child is not smaller than that at its left
sibling.  This left-dominance property is fulfilled for every right
child in a fine heap~\cite{CarlssonCM96} (see also
\cite{mdrsort,Weg92}), which uses one extra bit per node to maintain
the property.  By analogy, the heap property can be seen as a
dominance relation between a node and its parent.  As a binary heap, a
strong heap is viewed as a nearly complete binary tree where the
lowest level may be missing some nodes at the rightmost (last)
positions. Also, this tree is embedded in an array in the same way.
If the array indexing starts at $0$, the parent of a node at index $i$
($i \neq 0$) is at index $\lfloor (i-1)/2 \rfloor$, the left child (if
any) at index $2i+1$, and the right child (if any) at index $2i+2$.

\begin{figure}[b!]
\hspace*{1em}%
\begin{minipage}[b]{0.485\textwidth}
\begin{tikzpicture}[scale=0.71]
\tikzstyle{every path} = [draw,ultra thick]
\tikzstyle{every node} = [draw,shape=circle,very thick,minimum size=0.5cm,inner sep=0pt]



\node (4) at (1.0,1.5) {$4$};
\node (6) at (0.25,0.25) {$6$};
\node (15) at (1.75,0.25) {\raisebox{-0.8ex}[0pt][0pt]{\makebox[0pt]{$15$}}};

\draw[->] (4) -- (6);
\draw[->] (4) -- (15);
\draw[->] (6) -- (15);

\node (5) at (3.5,1.5) {$5$};
\node (7) at (2.75,0.25) {$7$};
\node (11) at (4.25,0.25) {$11$};

\draw[->] (5) -- (7);
\draw[->] (5) -- (11);
\draw[->] (7) -- (11);

\node (3) at (2.25,2.75) {$3$};
\draw[->] (3) -- (4);
\draw[->] (3) -- (5);
\draw[->] (4) -- (5);


\node (9) at (6.0,1.5) {$9$};
\node (10) at (5.25,0.25) {$10$};
\node (12) at (6.75,0.25) {$12$};

\draw[->] (9) -- (10);
\draw[->] (9) -- (12);
\draw[->] (10) -- (12);

\node (13) at (8.5,1.5) {$13$};
\node (14) at (7.75,0.25) {$14$};
\node (17) at (9.25,0.25) {$17$};

\draw[->] (13) -- (14);
\draw[->] (13) -- (17);
\draw[->] (14) -- (17);

\node (8) at (7.25,2.75) {$8$};
\draw[->] (8) -- (9);
\draw[->] (8) -- (13);
\draw[->] (9) -- (13);


\node (1) at (4.75,4) {$1$};
\draw[->] (1) -- (3);
\draw[->] (1) -- (8);
\draw[->] (3) -- (8);

\end{tikzpicture}
\end{minipage}\hspace*{3em}%
\begin{minipage}[t]{0.485\textwidth}
\begin{tikzpicture}[scale=0.71]
\tikzstyle{every path} = [draw,ultra thick]
\tikzstyle{every node} = [draw,shape=circle,very thick,minimum size=0.5cm,inner sep=0pt]


\node (1) at (4,7.75) {$1$};
\node (3) at (4,6.5) {$3$};
\node (4) at (1.5,5.25) {$4$};
\node (8) at (6.5,5.25) {$8$};
\node (6) at (0.25,4) {$6$};
\node (5) at (3.0,4) {$5$};
\node (9) at (6.5,4) {$9$};
\node (15) at (0.25,2.75){\raisebox{-0.8ex}[0pt][0pt]{\makebox[0pt]{$15$}}};
\node (7) at (3.0,2.75) {$7$};
\node (11) at (3.0,1.5) {$11$};
\node (10) at (5.25,2.75) {$10$};
\node (13) at (7.75,2.75) {$13$};
\node (12) at (5.25,1.5) {$12$};
\node (14) at (7.75,1.5) {$14$};
\node (17) at (7.75,0.25) {$17$};

\draw[->] (1) -- (3);
\draw[->] (3) -- (4);
\draw[->] (3) -- (8);
\draw[->] (4) -- (6);
\draw[->] (4) -- (5);
\draw[->] (8) -- (9);
\draw[->] (6) -- (15);
\draw[->] (5) -- (7);
\draw[->] (7) -- (11);
\draw[->] (9) -- (10);
\draw[->] (9) -- (13);
\draw[->] (10) -- (12);
\draw[->] (13) -- (14);
\draw[->] (14) -- (17);

\end{tikzpicture}
\end{minipage}
\caption{A strong heap in an array $a =
  [1,3,8,4,5,9,13,6,15,7,11,10,12,14,17]$ and its alternative
  representations: directed acyclic graph (left) and stretched tree
  (right)%
\label{fig2}}
\end{figure}
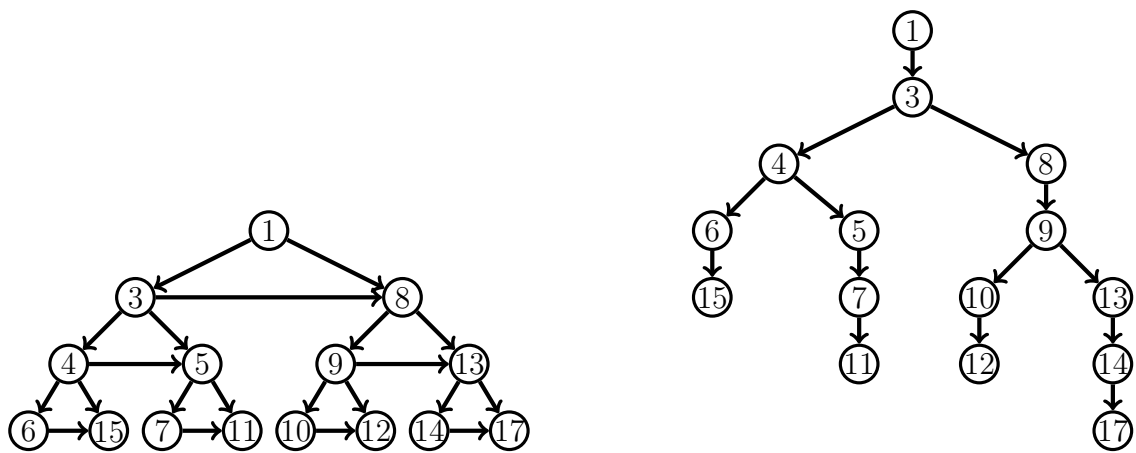

Two alternative representations of a strong heap are exemplified in
Fig.~\ref{fig2}.  On the left, the \emph{directed acyclic graph} has a
complete binary tree as its skeleton: There are arcs from every
parent to its children and additional arcs from every left child to its
sibling indicating the dominance relations.  On the right, in the
\emph{stretched tree} the arcs from each parent to its right child are
removed reflecting that these dominance relations can be induced.  In
the stretched tree a node can have 0, 1, or 2 children. A node has one
child if in the skeleton it is a right child that is not a leaf or it
is a leaf that has a right sibling. A node has two children if in the
skeleton it is a left child that is not a leaf.  If the underlying
nearly complete binary tree has height $h$, the height of the
stretched tree is at most $2h-1$ and on any path from the root to a leaf 
in the stretched tree the number of nodes with two children is at
most $h-2$.

The basic primitive used in the manipulation of a binary heap is the
\Siftdown{} \cite{Flo64,Wil64}. The procedure starts at a node that
breaks the heap property, traverses down the heap by following the
smaller children, and moves the encountered elements one level
up until the correct place of the element we started with is found. In
principle, \Siftdown{} performs a single iteration of
insertion sort to get the elements on this path in sorted
order. In a strong heap, the \Strengtheningsiftdown{} procedure does
the same, except that it operates on the stretched tree.
Having this tool available, \Extractmin{} can be implemented by
replacing the element at the root with the element at the last leaf
of the array, and invoking the \Strengtheningsiftdown{} procedure for
the root. 

\begin{example}
Consider the strong heap in Fig.~\ref{fig2}. If its minimum was
removed, the special path to be followed would include the nodes
$\langle$\circled{3}, \circled{4}, \circled{5}, \circled{7}, \circled{11}$\rangle$.
\end{example}

To build a strong heap for a given set of elements, we can imitate
Floyd's heap-construction algorithm \cite{Flo64}; that is,
\Strengtheningsiftdown{} is invoked for all nodes, one by one,
processing them in reverse order.

Let $n$ denote the size of the strong heap under consideration, and
let $h$ denote the height of the underlying nearly complete binary
tree.  When we go down the stretched tree, we have to perform at most
$h-2$ element comparisons (branching at binary nodes), and when we
check whether to stop or not, we have to perform at most $2h-1$
element comparisons.  Hence, the number of element comparisons
performed by \Extractmin{} is bounded by $3 h - 3$, which is at most
$3\lg n$ knowing that $h = \lfloor \lg n \rfloor + 1$. The total
number of element comparisons performed by \Construct{} is bounded by
the sum $\sum_{i=1}^{\lfloor \lg n \rfloor + 1} 3\cdot i \cdot \lceil
n/2^{i+1} \rceil$, which is at most $3n + o(n)$. For both procedures,
the amount of work done is proportional to the number of element
comparisons performed, i.e.~the worst-case running time of
\Extractmin{} is $O(\lg n)$ and that of \Construct{} is $O(n)$.

\begin{lemma}
A strong heap of size $n$ can be built in $O(n)$
worst-case time by repeatedly calling \Strengtheningsiftdown{}.  One
\Strengtheningsiftdown{} operation uses $O(\lg n)$ worst-case time and
performs at most $3\lg n$ element comparisons.
\end{lemma}

The next question is how to perform a \Siftdown{} operation on a
strong heap of size $n$ using at most $\lg n +O(1)$ element
comparisons.  At this stage we allow the amount of work to be much
higher, namely $O(n)$.  We show how to achieve these bounds when the
heap is \emph{complete}, i.e.~when all leaves have the same depth.

To keep the heap complete, assume that the element at the root of a
strong heap is to be replaced by a new element.  The
\Rotatingsiftdown{} procedure traverses the left spine of the root of the skeleton
bottom up starting from the leftmost leaf, and determines the correct
place of the new element using one element comparison at each node
visited.  Thereafter, it moves all the elements above this position on
the left spine one level up, and inserts the new element into this
place.  If this place is at level $h$, we have performed $h$ element
comparisons.  Up along the left spine of the root there are $\lg n - h
+ O(1)$ remaining levels to which we have moved other elements. While
this results in a heap, we still have to reinforce the left-dominance
property at these upper levels.  In accordance, we compare each
element that has moved up with its right sibling.  If the element at
position $j$ is larger than the element at position $j+1$, we have to
interchange the subtrees $T_j$ and $T_{j+1}$ rooted at positions $j$
and $j+1$ in-place by swapping all their elements. The
procedure continues this way until the root is processed.

\begin{example}
Consider the strong heap in Fig.~\ref{fig2}. If the element at the
root was replaced with element 16, the left spine to be followed would
include the nodes
$\langle$\circled{3}, \circled{4}, \circled{6}$\rangle$, the new
element would be placed at the leaf we ended up, the elements above
would be lifted up one level, and the subtree interchange would be
necessary for the subtrees rooted at node \circled{6} and its new
sibling \circled{5}.
\end{example}

Given two subtrees that are both complete and of height $h$, the
number of element moves needed to interchange the two subtrees is
$O(2^h)$.  As the sum of $O(2^h)$ over all $h=1,\ldots,\lfloor \lg n
\rfloor$ is $O(n)$, the total work done in the subtree interchanges is
$O(n)$.  Thus, the \Rotatingsiftdown{} operation requires at most $\lg
n + O(1)$ element comparisons and $O(n)$ work.

\begin{lemma}
In a complete strong heap of size $n$, a single \Rotatingsiftdown{}
operation can be executed in-place using at most $\lg n + O(1)$
element comparisons.  The number of element moves performed is $O(n)$.
\end{lemma}

\section{Lazy Heaps: Buffering Insertions}
\label{sec:buff}

In the variant of a binary heap that we describe in this section some
nodes may violate heap order, because insertions are buffered and
unordered bulks are incrementally melded with the heap.  The main difference
between the present construction and the construction in
\cite{EEK13} is that, for a heap of size $n$, we allow $O(\lg^2
n)$ heap-order violations, instead of $O(\lg n)$, but still we use
$O(1)$ extra space to track where the potential violations are.  By
replacing \Siftdown{} with \Strengtheningsiftdown{} the construction
will also work for strong heaps.

A \emph{lazy heap} is composed of three parts: 1)~\emph{main
  heap},  2)~\emph{submersion area}, and 3)~\emph{insertion buffer}. 
  The following rules are imposed:

\begin{enumerate}
\item[1)] New insertions are appended to the insertion buffer at the end of the array. 
\item[2)] The submersion area is incrementally melded with the main heap
  by performing a constant amount of work in connection with every
  \Insert{} operation.
\item[3)] When the insertion buffer becomes full, the submersion process must have been completed. A proportion of the elements of the insertion buffer are then treated as a new submersion area.
\item[4)] When the insertion buffer is empty, the submersion process must have been completed.
When an \Extractmin{} operation is performed, a replacement element is thus borrowed from either the insertion buffer or the main heap 
(if the insertion buffer is empty) but never from the submersion area.
\end{enumerate}

The insertion buffer occupies the last locations of the array.  If the
size of the main heap is $n_0$, the insertion buffer is never allowed
to become larger than $\Theta(\lg^2 n_0)$.  The buffer should support
insertions in constant time, minimum extractions in $O(\lg n)$ time
using at most $\lg n + O(1)$ element comparisons, and it should be
possible to meld a buffer with the main heap efficiently.

Let $t = \lfloor \lg(1+ \lg(1+n_0)) \rfloor$.  We treat the insertion
buffer as a sequence of \emph{chunks}, each of size $k= 2^{t-1}$, and
we limit the number of chunks to at most~$k$.  All the chunks, except
possibly the last one, will contain exactly $k$ elements.  The minimum
of each chunk is kept at the first location of the chunk, and the
index of the minimum of the buffer is maintained.  When this minimum is
removed, the last element is moved into its place, the new minimum of
that chunk is found in $O(k)$ time using $k-1$ element comparisons (by
scanning the elements of the chunk), and then the new overall minimum
of the buffer is found in $O(k)$ time using $k-1$ element comparisons
(by scanning the minima of the chunks).

When an \Extractmin{} operation needs a replacement for the old minimum, we
have to consider the case where the last element is the overall minimum
of the insertion buffer.  In such a case, to avoid losing track of
this minimum, we start by swapping the last element of the insertion
buffer with its first element.

In \Insert{}, a new element is appended to the buffer.  
Subsequently, the minimum of the last
chunk and the minimum of the buffer are adjusted if necessary;
this requires at most two element comparisons.  Once there are $k$
full chunks, the first half of them are used to form a new submersion area and
the elements therein are incrementally inserted into the main heap as a bulk.

A \Bulkinsert{} procedure is performed incrementally by visiting the
ancestors of the nodes containing the initial bulk as in Floyd's
heap-construction algorithm \cite{Flo64}.  During such a submersion,
the submersion area is treated as part of the main heap whose nodes
may not obey the heap property.  To perform a bulk insertion, heap
order is reestablished bottom up level by level.  Starting with the
parents of the nodes containing the initial bulk, for each node we
call the \Siftdown{} subroutine. We then consider the parents of these
nodes at the next upper level, restoring heap order up to this
level. This process is repeated all the way up to the root.  As long
as there are more than two nodes that are considered at a level, the
number of such nodes almost halves at the next level.

In the following analysis we separately consider two phases of the
\Bulkinsert{} procedure.  The first phase comprises the \Siftdown{}
calls for the nodes at the levels with more than two involved nodes.
Let $b$ denote the size of the initial bulk.  The number of the nodes
visited at the $j$th last level is at most $\lfloor(b-2)/2^{j-1}
\rfloor + 2$. For a node at the $j$th last level, a call to the
\Siftdown{} subroutine requires $O(j)$ work. In the first phase the
amount of work involved is $O(\sum_{j=2}^{\lceil \lg r \rceil}
j/2^{j-1} \cdot b) = O(b)$.  The second phase comprises at most
$2\lfloor \lg n_0 \rfloor$ calls to the \Siftdown{} subroutine; this
accounts for a total of $O(\lg^2 n_0)$ work.  Since $b = \Theta(\lg^2
n_0)$, the overall work done is $O(\lg^2 n_0)$, i.e.~amortized
constant per \Insert{}.  Instead of doing this job in one shot, we
distribute the work by performing $O(1)$ work in connection with every
heap operation.  Obviously, the bulk insertion should be done fast
enough so that it completes before the insertion buffer becomes either
full or empty.

To track the progress of the submersion process, we maintain
two intervals that represent the nodes up to which the \Siftdown{}
subroutine has been called. Each such interval is represented by two
indices indicating its left and right endpoints, call them $(\ell_1,r_1)$
and $(\ell_2,r_2)$. These two intervals are at two consecutive levels, 
and the parent of the right endpoint of the first interval has an index that is one less than the left endpoint
of the second interval, i.e. $\ell_2 - 1 = \lfloor (r_1-1)/2 \rfloor$.  
We call these two intervals the \emph{frontier}. 
Notice that the submersion area consists of all the descendants of the nodes at the frontier. 
While the submersion process advances, the frontier moves upwards and shrinks
until it has one or two nodes.  This frontier imparts that a
\Siftdown{} is being performed starting from the node whose index is
$\ell_2 - 1$.  In addition to the frontier, we also maintain the index of
the node that the \Siftdown{} in progress is currently processing.
In connection with every heap operation, the current \Siftdown{}
progresses a constant number of levels downwards and this index is
updated. Once the \Siftdown{} operation returns, the frontier is
updated. When the frontier reaches the root, the submersion process is
complete.  To summarize, the information maintained to record the
state of a bulk insertion is two intervals of indices to represent
the frontier plus the node which is under consideration by the current
\Siftdown{} operation.

As for the insertion buffer, we also maintain the index of the minimum
among the elements on the frontier.  We treat each of the two
intervals of the frontier as a set of consecutive chunks.  Except for
the first or last chunk on each interval that may have less nodes,
every other chunk has $k$ nodes.  In addition, we maintain the
invariant that the minimum within every chunk on the frontier is kept
at the entry storing the first node among the nodes of the chunk.  An
exception is the first and last chunks, where we maintain the index
for the minimum on each.

To remove the minimum of the submersion buffer, we know it must be on
the frontier and readily have its index.  This minimum is swapped with
the last element of the array and a \Siftdown{} is performed to remedy
the order between the replacement element and the elements in its descendants.  
We distinguish between two cases: 1) If there are at most two nodes on
the frontier, we make the minimum index of the frontier point to the
smaller.  2) If there are more than two nodes on the frontier, the
height of the nodes on the frontier is at most $2\lg\lg n + O(1)$.  The
nodes of the chunk that contained the removed minimum are scanned to
find its new minimum.  If this chunk is neither the first nor the last
of the frontier, its new minimum is swapped with the element at its
first position, and this is followed by a \Siftdown{} performed on the
latter element.  The overall minimum of the frontier is then localized
by scanning the minima of all the chunks.
Extracting the minimum of the submersion buffer thus requires 
$O(\lg n)$ time using at most $\lg n +O(1)$ element comparisons.

\section{Strengthened Lazy Heaps: Putting Things Together}
\label{sec:ultimate}

Our final construction is similar to the one of the previous section in that it
consists of three components: main heap, submersion area, and insertion buffer. 
The main heap has two layers: a \emph{top heap} that is a binary heap,
and each leaf of the top heap roots a \emph{bottom heap} that is a
complete strong heap. Because 
the main heap is only partially strong, we call the resulting data
structure a \emph{strengthened lazy heap}.

Let $n_0$ be the size of the main heap, and let $t=\lfloor \lg(1+ \lg
(1+n_0)) \rfloor$.  The height of the bottom heaps is $t$, or possibly
$t+1$. In the insertion buffer, the size of a chuck is $k=2^{t-1}$ and
the size of the buffer is bounded by $k^2$.

\begin{figure}[tb!]
\hspace*{1.5cm}\begin{tikzpicture}[scale=0.71]


\path[draw,very thick] (0.5,2.0) -- (4.18,7.0) -- (7.5,2.5);
\path[draw,very thick] (0.5,2.0) -- (2,2) -- (2,2.5) -- (7.5,2.5);

\path[draw,very thick] (0,0) -- (0.5,2) -- (1,0) -- cycle;
\path[draw,very thick] (1,0) -- (1.5,2) -- (2,0) -- cycle;

\path[draw,very thick] (2,0) -- (2.5,2.5) -- (3,0) -- cycle;
\path[draw,very thick] (3,0) -- (3.5,2.5) -- (4,0) -- cycle;

\path[draw,very thick] (4,0) -- (4.5,2.5) -- (5,0) -- cycle;
\path[draw,very thick] (5,0) -- (5.5,2.5) -- (6,0) -- cycle;

\path[draw,very thick] (6,0.5) -- (6.5,2.5) -- (7,0.5) -- cycle;
\path[draw,very thick] (7,0.5) -- (7.5,2.5) -- (8,0.5) -- cycle;

\path[draw,line width=0pt,fill=gray!50] (4.1,0.0) rectangle (4.9,0.5);
\path[draw,line width=0pt,fill=gray!50] (5.2,0.5) rectangle (5.4,1.0);
\path[draw,line width=0pt,fill=gray!50] (5.4,1.0) rectangle (5.7,1.5);
\node[line width=0pt,fill=gray!50] (frontier) at (-0.9, 0.75) {\small frontier};


\path[draw,very thick] (6.1,-0.1) rectangle (7.7,0.4);

\node[line width=0pt] (t) at (-0.5, 2) {\small $t$};
\node[line width=0pt] (top) at (8, 5) {\small \makebox[0cm][l]{top heap}};
\node[line width=0pt] (top) at (8, 2.5) {\small \makebox[0cm][l]{border}};
\node[line width=0pt] (bottom) at (8, 1.25) {\small \makebox[0cm][l]{bottom heaps}};
\node[line width=0pt] (area) at (6.5, -0.7) {\small
  \makebox[0cm][r]{submersion area}};
\node[line width=0pt] (buffer) at (8, 0.2) {\small \makebox[0cm][l]{insertion buffer}};

\shade[shading=ball, ball color=black] (0.5,2) circle (0.125);
\shade[shading=ball, ball color=black] (1.5,2) circle (0.125);
\shade[shading=ball, ball color=black] (2.5,2.5) circle (0.125);
\shade[shading=ball, ball color=black] (3.5,2.5) circle (0.125);
\shade[shading=ball, ball color=black] (4.5,2.5) circle (0.125);
\shade[shading=ball, ball color=black] (5.5,2.5) circle (0.125);
\shade[shading=ball, ball color=black] (6.5,2.5) circle (0.125);
\shade[shading=ball, ball color=black] (7.5,2.5) circle (0.125);

\end{tikzpicture}
\caption{Schematic view of a strengthened lazy heap\label{structure}}
\end{figure}
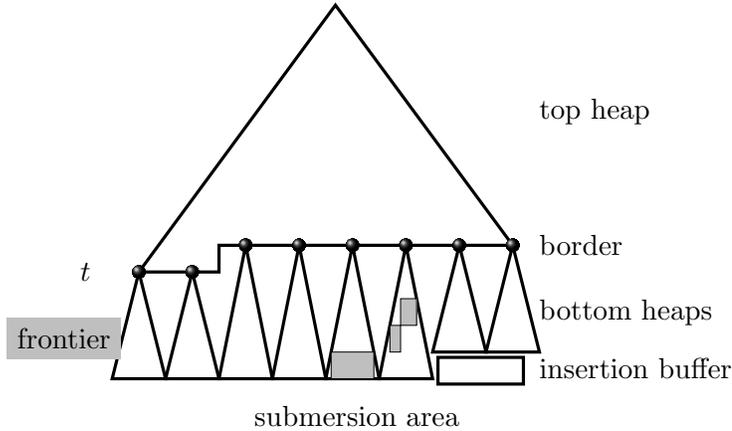

Because of the two-layer structure, the incremental remedy processes
are more complicated for a strengthened lazy heap than for a lazy
heap. Let us consider the introduced complications one at a
time. To help the reader to get a complete picture of the data
structure, we visualize it in Fig.~\ref{structure}.

\emph{Complication} 1. Due to insertions and extractions, we should be
able to move the \emph{border} between the two layers dynamically.  To
make the bottom heaps one level shallower, we can just adjust $t$ and
ignore the left domination for the children of the nodes at the
previous border. To make the bottom heaps one level higher, we need a
new incremental remedy process that scans the nodes on the old border
and compares every left child with its right sibling.  If the right
sibling is smaller, the two elements are swapped and a
\Strengtheningsiftdown{} operation is incrementally applied on the right
sibling.  Again, we only need a constant amount of space to record the
current state of this process. When moving the border upwards,
the total work done is linear so, after the process is initiated,
every forthcoming operation has to take a constant share of this work. 
The \Extractmin{} operations do not need to be aware of
this process, as they can always make a conservative
assumption about the level at which the border is. 

\emph{Complication} 2. We need a new procedure, 
that we call \Combinedsiftdown{}, instead of \Siftdown{}.  Assume we have to replace the minimum 
element that is in the top heap with another element.  To reestablish the heap properties, we
follow the proposal of Carlsson~\cite{Car87}: We traverse down 
along the path of the smaller children until we reach a root of a
bottom heap. By comparing the replacement element with the element at the root
of this bottom heap, we check whether the replacement element should land in
the top heap or in the bottom heap. In the first case, we find the
position of the replacement element using binary search on the traversed path;
the path is traversed sequentially, but element comparisons are made
in a binary-search manner.  In the second case, we apply the
\Rotatingsiftdown{} operation for the root of the bottom heap.

\emph{Complication} 3. Normally, we use the last element in
the array as a replacement for the old minimum.  However,
if the insertion buffer is empty, meaning that the submersion
process must have been completed, we need to use an element from the
main heap as a replacement.  In such a case, to keep the bottom heaps
complete, our solution is to move all the elements at the lowest level
of the last bottom heap (these are the last elements of the array)
back to the empty insertion buffer.  We consciously adjusted the
parameters such that the number of elements in one chunk of the
insertion buffer matches the number of elements at the lowest
level of the last bottom heap.  After such a move, the minimum of this
chunk is not known.  Fortunately, we would never need this minimum
within the next $k$ \Extractmin{} operations, as there are at least a
logarithmic number of elements in the main heap that are smaller.  In
such a case, the minimum of this chunk is incrementally found within
the upcoming $k$ heap operations.

\emph{Complication} 4. In the top heap the frontier causes no problem and is treated as before. 
However, if we perform a \Rotatingsiftdown{} in a bottom heap
whose frontier consists of two intervals, there is a risk that we mess
up the frontier. In accordance, we schedule the submersion process
differently: We process the bottom heaps under submersion one by one,
and lock the bottom heap that is under consideration while performing 
\Extractmin{} operations. When the
frontier overlaps the bottom heaps, it is cut into several pieces:
1)~the interval corresponding to the unprocessed leaves of the initial
bulk, 2)~the two intervals $(\ell_1,r_1)$ and $(\ell_2,r_2)$ in the
bottom heap under consideration, and 3)~the interval of the roots of
the bottom heaps that have been handled by the submersion process.
Above the unprocessed interval and below the processed interval,
\Rotatingsiftdown{} operations can be performed without 
complications. Locking solves the problem for the bottom heap under
consideration.  However, in that bottom heap there are some nodes
between the root and the frontier that are in no man's land since they
are locked, but not yet included in the submersion process.  These
nodes will accordingly not be in order with the elements above or
below.  This is not a problem however, as none of these elements can be the
minimum of the heap except after a logarithmic number of operations.  Within such
time, these nodes should have already been handled by the submersion
process.

\emph{Complication} 5. There should not be any interference between
the remedy processes.  One issue arises when a \Rotatingsiftdown{}
operation meets a node that an incremental \Strengtheningsiftdown{}
operation is processing. To avoid wrong decisions, it only requires
one more check in the \Rotatingsiftdown{} procedure if it meets such a
node on the left spine of a bottom heap.

Let us now recap how the operations are executed and analyse their
performance. Here we ignore the extra work done due to the incremental
processes. Clearly, \Minimum{} can be carried out in $O(1)$
worst-case time by reporting the minimum of three elements: 1)~the
element at the root of the top heap, 2)~the minimum of the
insertion buffer, and 3)~the minimum of the submersion area.  As
before, \Insert{} appends the given element to the insertion buffer
and updates the minimum of the buffer if necessary.  To perform an
\Extractmin{} operation, we need to consider the different minima and
remove the smallest among them.

\emph{Case} 1. If the minimum is at the root of the top heap, we find
a replacement for the old minimum and apply the \Combinedsiftdown{}
for the root. If we meet the frontier, we stop the operation before
crossing it.  Let $n$ denote the total number of elements. The top
heap is of size $O(n/\lg n)$ and the bottom heaps are of size $O(\lg
n)$.  To reach the root of a bottom heap, we perform $\lg n - \lg\lg n
+ O(1)$ element comparisons. If we have to go upwards, we perform
$\lg\lg n + O(1)$ additional element comparisons in the binary
search. On the other hand, if we have to go downwards, the
\Rotatingsiftdown{} operation needs to perform at most $\lg\lg n +
O(1)$ element comparisons. In both cases, the work done is
$O(\lg n)$.

\emph{Case} 2. If the minimum of the insertion buffer
is the overall minimum, it is removed as explained in the previous
section.  The operation involves $2k + O(1)$ element comparisons and
the amount of work done is proportional to that number. 
Since we have set $k = 2^{t-1} = \frac{1}{2} \lg n +O(1)$, 
the minimum extraction here also requires at most $\lg n + O(1)$ element
comparisons and $O(\lg n)$ work.

\emph{Case} 3. If the frontier contains the overall minimum, we apply
a similar treatment to that explained in the previous section with a basic exception.
If there are more than two nodes on the frontier, the height of the nodes on the frontier is at most $2 \lg \lg n +O(1)$. 
In this case, we use the \Strengtheningsiftdown{} procedure in place of the \Siftdown{} procedure. 
If there are at most two nodes on the frontier, The frontier lies in the top heap.
In this case, we apply the \Combinedsiftdown{} procedure instead.
Either way, the minimum extraction requires at most $\lg n + O(1)$ element
comparisons and $O(\lg n)$ work as well.

When executing the subtree interchanges, the number of element moves
performed---even though asymptotically logarithmic---would still be
larger than the number of element comparisons.  Alternatively, one can
control the number of element moves performed by the
\Rotatingsiftdown{} operation by adjusting the heights of the bottom
heaps.  If the maximum height of a bottom heap is set to
$t-\lg(3/\varepsilon)$, for some small constant $0 < \varepsilon \leq
3$, the number of element moves performed by the \Rotatingsiftdown{}
operation will be bounded by $\varepsilon \lg n + O(1)$, while the
aforementioned bounds for the other operations still hold.

\section{Conclusions}

We described a priority queue that 1)~operates in-place, 2)~supports
\Minimum{} and \Insert{} in $O(1)$ worst-case time, 3)~supports
\Extractmin{} in $O(\lg n)$ worst-case time, and 4)~performs at most
$\lg n + O(1)$ element comparisons per \Extractmin{}. The data
structure is asymptotically optimal with respect to time, and optimal
up-to-additive-constant with respect to space and element comparisons.

The developed data structure is a consequence of many years' work.  Previous
breakthroughs can be summarized as follows: 1)~break the $2\lg n +
O(1)$ barrier for the number of element comparisons performed per
\Extractmin{} when \Insert{} takes $O(1)$ worst-case time
\cite{Elm04}, 2)~achieve the desired bounds using $O(n)$ words of
extra space
\cite{EJK08}, 3)~reduce the amount of extra space to $O(n)$ bits
\cite{EEK13}, 4)~achieve the desired bounds in the amortized sense
\cite{EK13}.  And, in this paper, 
we developed an in-place structure that achieves the desired bounds in the worst case.

It is interesting that we could surpass the two lower bounds known for
binary heaps \cite{GM86} by slightly loosening the assumptions that
are intrinsic to these lower bounds.  To achieve our goals, we
simultaneously imposed more order on some nodes by forbidding some
left children to be larger than their right siblings and less order
on others by allowing some nodes to possibly be smaller than their
parents. Our last one-line tip is that it may not be necessary to maintain the heap
property---exactly as defined for binary heaps---after every operation!

%
\bibliographystyle{splncs03}
\bibliography{shortstrings,icalp-14}

\begin{thebibliography}{10}
\providecommand{\url}[1]{\texttt{#1}}
\providecommand{\urlprefix}{URL }

\bibitem{Bro78}
Brown, M.R.: Implementation and analysis of binomial queue algorithms. SIAM J.
  Comput.  7(3),  298--319 (1978)

\bibitem{Car87}
Carlsson, S.: A variant of {H}eapsort with almost optimal number of
  comparisons. Inform. Process. Lett.  24(4),  247--250 (1987)

\bibitem{Car91}
Carlsson, S.: An optimal algorithm for deleting the root of a heap. Inform.
  Process. Lett.  37(2),  117--120 (1991)

\bibitem{CarlssonCM96}
Carlsson, S., Chen, J., Mattsson, C.: Heaps with bits. Theoret. Comput. Sci.
  164(1--2),  1--12 (1996)

\bibitem{CMP88}
Carlsson, S., Munro, J.I., Poblete, P.V.: An implicit binomial queue with
  constant insertion time. In: Karlsson, R., Lingas, A. (eds.) SWAT 1988. LNCS,
  vol. 318, pp. 1--13. Springer, Heidelberg (1988)

\bibitem{CEEK12}
Chen, J., Edelkamp, S., Elmasry, A., Katajainen, J.: In-place heap construction
  with optimized comparisons, moves, and cache misses. In: Rovan, B., Sassone,
  V., Widmayer, P. (eds.) MFCS 2012. LNCS, vol. 7464, pp. 259--270. Springer,
  Heidelberg (2012)

\bibitem{Dut93}
Dutton, R.D.: Weak-heap sort. BIT  33(3),  372--381 (1993)

\bibitem{EEK13}
Edelkamp, S., Elmasry, A., Katajainen, J.: Weak heaps engineered. J. Discrete
  Algorithms  23,  83--97 (2013), {P}resented at IWOCA 2012

\bibitem{Elm04}
Elmasry, A.: Layered heaps. In: Hagerup, T., Katajainen, J. (eds.) SWAT 2004.
  LNCS, vol. 3111, pp. 212--222. Springer, Heidelberg (2004)

\bibitem{EJK08}
Elmasry, A., Jensen, C., Katajainen, J.: Multipartite priority queues. ACM
  Trans. Algorithms  5(1),  Article 14 (2008)

\bibitem{EK13}
Elmasry, A., Katajainen, J.: Towards ultimate binary heaps. CPH STL Report
  2013-1, Department of Computer Science, University of Copenhagen (2013)

\bibitem{Flo64}
Floyd, R.W.: Algorithm 245: {T}reesort 3. Commun. ACM  7(12),  701 (1964)

\bibitem{GM86}
Gonnet, G.H., Munro, J.I.: Heaps on heaps. SIAM J. Comput.  15(4),  964--971
  (1986), {P}resented at ICALP 1982

\bibitem{ultimate}
Katajainen, J.: The ultimate heapsort. In: CATS 1998. Australian Computer
  Science Communications, vol.~20, pp. 87--95. Springer, Singapore (1998)

\bibitem{Knu73}
Knuth, D.E.: Sorting and Searching, The Art of Computer Programming, vol.~3.
  Addison Wesley Longman, Reading, 2nd edn. (1998)

\bibitem{mdrsort}
McDiarmid, C.J.H., Reed, B.A.: Building heaps fast. J. Algorithms  10(3),
  352--365 (1989)

\bibitem{Vui78}
Vuillemin, J.: A data structure for manipulating priority queues. Commun. ACM
  21(4),  309--315 (1978)

\bibitem{Weg92}
Wegener, I.: The worst case complexity of {M}c{D}iarmid and {R}eed's variant of
  {B}ottom-up {H}eapsort is less than $n\log n + 1.1n$. Inform. and Comput.
  97(1),  86--96 (1992), {P}resented at STACS 1991

\bibitem{Wil64}
Williams, J.W.J.: Algorithm 232: {H}eapsort. Commun. ACM  7(6),  347--348
  (1964)

\end{thebibliography}
\end{document}